\documentclass[aps,preprint,onecolumn,amsmath,amssymb,nofootinbib,superscriptaddress]{revtex4-1}
\usepackage{times}
\usepackage[pdftex]{graphicx}
\usepackage{dcolumn}
\usepackage{bm}
\usepackage{amsmath}
\usepackage{indentfirst}
\usepackage{float}
\usepackage[colorlinks]{hyperref}
\usepackage[dvipsnames]{xcolor}
\usepackage{xcolor}
\usepackage{subfigure}
\usepackage{algorithm}
\usepackage{algorithmicx}
\usepackage{algpseudocode}
\usepackage{amsmath}
\usepackage{verbatim}

\begin{document}
	
	\title{Realization of fast all-microwave CZ gates with a tunable coupler }
	\author{Shaowei Li}
	\affiliation{Department of Modern Physics, University of Science and Technology of China, Hefei 230026, China}
	\affiliation{Shanghai Branch, CAS Center for Excellence in Quantum Information and Quantum Physics, University of Science and Technology of China, Shanghai 201315, China}
	\affiliation{Shanghai Research Center for Quantum Sciences, Shanghai 201315, China}
	
	\author{Daojin Fan}
	\affiliation{Department of Modern Physics, University of Science and Technology of China, Hefei 230026, China}
	\affiliation{Shanghai Branch, CAS Center for Excellence in Quantum Information and Quantum Physics, University of Science and Technology of China, Shanghai 201315, China}
	\affiliation{Shanghai Research Center for Quantum Sciences, Shanghai 201315, China}
	
	\author{Ming Gong}
	\affiliation{Department of Modern Physics, University of Science and Technology of China, Hefei 230026, China}
	\affiliation{Shanghai Branch, CAS Center for Excellence in Quantum Information and Quantum Physics, University of Science and Technology of China, Shanghai 201315, China}
	\affiliation{Shanghai Research Center for Quantum Sciences, Shanghai 201315, China}

	\author{Yangsen Ye}
	\affiliation{Department of Modern Physics, University of Science and Technology of China, Hefei 230026, China}
	\affiliation{Shanghai Branch, CAS Center for Excellence in Quantum Information and Quantum Physics, University of Science and Technology of China, Shanghai 201315, China}
	\affiliation{Shanghai Research Center for Quantum Sciences, Shanghai 201315, China}
	
	\author{Xiawei Chen}
	\affiliation{Department of Modern Physics, University of Science and Technology of China, Hefei 230026, China}
	\affiliation{Shanghai Branch, CAS Center for Excellence in Quantum Information and Quantum Physics, University of Science and Technology of China, Shanghai 201315, China}
	\affiliation{Shanghai Research Center for Quantum Sciences, Shanghai 201315, China}
	
	\author{Yulin Wu}
	\affiliation{Department of Modern Physics, University of Science and Technology of China, Hefei 230026, China}
	\affiliation{Shanghai Branch, CAS Center for Excellence in Quantum Information and Quantum Physics, University of Science and Technology of China, Shanghai 201315, China}
	\affiliation{Shanghai Research Center for Quantum Sciences, Shanghai 201315, China}
	
	\author{Huijie Guan}
	\affiliation{Department of Modern Physics, University of Science and Technology of China, Hefei 230026, China}
	\affiliation{Shanghai Branch, CAS Center for Excellence in Quantum Information and Quantum Physics, University of Science and Technology of China, Shanghai 201315, China}
	\affiliation{Shanghai Research Center for Quantum Sciences, Shanghai 201315, China}
	
	\author{Hui Deng}
	\affiliation{Department of Modern Physics, University of Science and Technology of China, Hefei 230026, China}
	\affiliation{Shanghai Branch, CAS Center for Excellence in Quantum Information and Quantum Physics, University of Science and Technology of China, Shanghai 201315, China}
	\affiliation{Shanghai Research Center for Quantum Sciences, Shanghai 201315, China}
	
	\author{Hao Rong}
	\affiliation{Department of Modern Physics, University of Science and Technology of China, Hefei 230026, China}
	\affiliation{Shanghai Branch, CAS Center for Excellence in Quantum Information and Quantum Physics, University of Science and Technology of China, Shanghai 201315, China}
	\affiliation{Shanghai Research Center for Quantum Sciences, Shanghai 201315, China}
	
	\author{He-Liang Huang}
	\affiliation{Department of Modern Physics, University of Science and Technology of China, Hefei 230026, China}
	\affiliation{Shanghai Branch, CAS Center for Excellence in Quantum Information and Quantum Physics, University of Science and Technology of China, Shanghai 201315, China}
	\affiliation{Shanghai Research Center for Quantum Sciences, Shanghai 201315, China}

	\author{Chen Zha}
	\affiliation{Department of Modern Physics, University of Science and Technology of China, Hefei 230026, China}
	\affiliation{Shanghai Branch, CAS Center for Excellence in Quantum Information and Quantum Physics, University of Science and Technology of China, Shanghai 201315, China}
	\affiliation{Shanghai Research Center for Quantum Sciences, Shanghai 201315, China}
	
	\author{Kai Yan}
	\affiliation{Department of Modern Physics, University of Science and Technology of China, Hefei 230026, China}
	\affiliation{Shanghai Branch, CAS Center for Excellence in Quantum Information and Quantum Physics, University of Science and Technology of China, Shanghai 201315, China}
	\affiliation{Shanghai Research Center for Quantum Sciences, Shanghai 201315, China}
	
	\author{Shaojun Guo}
	\affiliation{Department of Modern Physics, University of Science and Technology of China, Hefei 230026, China}
	\affiliation{Shanghai Branch, CAS Center for Excellence in Quantum Information and Quantum Physics, University of Science and Technology of China, Shanghai 201315, China}
	\affiliation{Shanghai Research Center for Quantum Sciences, Shanghai 201315, China}
	
	\author{Haoran Qian}
	\affiliation{Department of Modern Physics, University of Science and Technology of China, Hefei 230026, China}
	\affiliation{Shanghai Branch, CAS Center for Excellence in Quantum Information and Quantum Physics, University of Science and Technology of China, Shanghai 201315, China}
	\affiliation{Shanghai Research Center for Quantum Sciences, Shanghai 201315, China}
	
	\author{Haibin Zhang}
	\affiliation{Department of Modern Physics, University of Science and Technology of China, Hefei 230026, China}
	\affiliation{Shanghai Branch, CAS Center for Excellence in Quantum Information and Quantum Physics, University of Science and Technology of China, Shanghai 201315, China}
	\affiliation{Shanghai Research Center for Quantum Sciences, Shanghai 201315, China}
	
	\author{Fusheng Chen}
	\affiliation{Department of Modern Physics, University of Science and Technology of China, Hefei 230026, China}
	\affiliation{Shanghai Branch, CAS Center for Excellence in Quantum Information and Quantum Physics, University of Science and Technology of China, Shanghai 201315, China}
	\affiliation{Shanghai Research Center for Quantum Sciences, Shanghai 201315, China}
	
	\author{Qingling Zhu}
	\affiliation{Department of Modern Physics, University of Science and Technology of China, Hefei 230026, China}
	\affiliation{Shanghai Branch, CAS Center for Excellence in Quantum Information and Quantum Physics, University of Science and Technology of China, Shanghai 201315, China}
	\affiliation{Shanghai Research Center for Quantum Sciences, Shanghai 201315, China}
	
	\author{Youwei Zhao}
	\affiliation{Department of Modern Physics, University of Science and Technology of China, Hefei 230026, China}
	\affiliation{Shanghai Branch, CAS Center for Excellence in Quantum Information and Quantum Physics, University of Science and Technology of China, Shanghai 201315, China}
	\affiliation{Shanghai Research Center for Quantum Sciences, Shanghai 201315, China}
	
	\author{Shiyu Wang}
	\affiliation{Department of Modern Physics, University of Science and Technology of China, Hefei 230026, China}
	\affiliation{Shanghai Branch, CAS Center for Excellence in Quantum Information and Quantum Physics, University of Science and Technology of China, Shanghai 201315, China}
	\affiliation{Shanghai Research Center for Quantum Sciences, Shanghai 201315, China}
	
	\author{Chong Ying}
	\affiliation{Department of Modern Physics, University of Science and Technology of China, Hefei 230026, China}
	\affiliation{Shanghai Branch, CAS Center for Excellence in Quantum Information and Quantum Physics, University of Science and Technology of China, Shanghai 201315, China}
	\affiliation{Shanghai Research Center for Quantum Sciences, Shanghai 201315, China}
	
	\author{Sirui Cao}
	\affiliation{Department of Modern Physics, University of Science and Technology of China, Hefei 230026, China}
	\affiliation{Shanghai Branch, CAS Center for Excellence in Quantum Information and Quantum Physics, University of Science and Technology of China, Shanghai 201315, China}
	\affiliation{Shanghai Research Center for Quantum Sciences, Shanghai 201315, China}
	
	\author{Jiale Yu}
	\affiliation{Department of Modern Physics, University of Science and Technology of China, Hefei 230026, China}
	\affiliation{Shanghai Branch, CAS Center for Excellence in Quantum Information and Quantum Physics, University of Science and Technology of China, Shanghai 201315, China}
	\affiliation{Shanghai Research Center for Quantum Sciences, Shanghai 201315, China}
	
	\author{Futian Liang}
	\affiliation{Department of Modern Physics, University of Science and Technology of China, Hefei 230026, China}
	\affiliation{Shanghai Branch, CAS Center for Excellence in Quantum Information and Quantum Physics, University of Science and Technology of China, Shanghai 201315, China}
	\affiliation{Shanghai Research Center for Quantum Sciences, Shanghai 201315, China}
	
	\author{Yu Xu}
	\affiliation{Department of Modern Physics, University of Science and Technology of China, Hefei 230026, China}
	\affiliation{Shanghai Branch, CAS Center for Excellence in Quantum Information and Quantum Physics, University of Science and Technology of China, Shanghai 201315, China}
	\affiliation{Shanghai Research Center for Quantum Sciences, Shanghai 201315, China}
	
	\author{Jin Lin}
	\affiliation{Department of Modern Physics, University of Science and Technology of China, Hefei 230026, China}
	\affiliation{Shanghai Branch, CAS Center for Excellence in Quantum Information and Quantum Physics, University of Science and Technology of China, Shanghai 201315, China}
	\affiliation{Shanghai Research Center for Quantum Sciences, Shanghai 201315, China}
	
	\author{Cheng Guo}
	\affiliation{Department of Modern Physics, University of Science and Technology of China, Hefei 230026, China}
	\affiliation{Shanghai Branch, CAS Center for Excellence in Quantum Information and Quantum Physics, University of Science and Technology of China, Shanghai 201315, China}
	\affiliation{Shanghai Research Center for Quantum Sciences, Shanghai 201315, China}
	
	\author{Lihua Sun}
	\affiliation{Department of Modern Physics, University of Science and Technology of China, Hefei 230026, China}
	\affiliation{Shanghai Branch, CAS Center for Excellence in Quantum Information and Quantum Physics, University of Science and Technology of China, Shanghai 201315, China}
	\affiliation{Shanghai Research Center for Quantum Sciences, Shanghai 201315, China}
	
	\author{Na Li}
	\affiliation{Department of Modern Physics, University of Science and Technology of China, Hefei 230026, China}
	\affiliation{Shanghai Branch, CAS Center for Excellence in Quantum Information and Quantum Physics, University of Science and Technology of China, Shanghai 201315, China}
	\affiliation{Shanghai Research Center for Quantum Sciences, Shanghai 201315, China}
	
	\author{Lianchen Han}
	\affiliation{Department of Modern Physics, University of Science and Technology of China, Hefei 230026, China}
	\affiliation{Shanghai Branch, CAS Center for Excellence in Quantum Information and Quantum Physics, University of Science and Technology of China, Shanghai 201315, China}
	\affiliation{Shanghai Research Center for Quantum Sciences, Shanghai 201315, China}
	
	\author{Cheng-Zhi Peng}
	\affiliation{Department of Modern Physics, University of Science and Technology of China, Hefei 230026, China}
	\affiliation{Shanghai Branch, CAS Center for Excellence in Quantum Information and Quantum Physics, University of Science and Technology of China, Shanghai 201315, China}
	\affiliation{Shanghai Research Center for Quantum Sciences, Shanghai 201315, China}
	
	\author{Xiaobo Zhu}
	\email{xbzhu16@ustc.edu.cn}
	\affiliation{Department of Modern Physics, University of Science and Technology of China, Hefei 230026, China}
	\affiliation{Shanghai Branch, CAS Center for Excellence in Quantum Information and Quantum Physics, University of Science and Technology of China, Shanghai 201315, China}
	\affiliation{Shanghai Research Center for Quantum Sciences, Shanghai 201315, China}
	
	\author{Jian-Wei Pan}
	\affiliation{Department of Modern Physics, University of Science and Technology of China, Hefei 230026, China}
	\affiliation{Shanghai Branch, CAS Center for Excellence in Quantum Information and Quantum Physics, University of Science and Technology of China, Shanghai 201315, China}
	\affiliation{Shanghai Research Center for Quantum Sciences, Shanghai 201315, China}
	
	\date{\today}
	
	
	\begin{abstract}
		The development of high-fidelity two-qubit quantum gates is essential for digital quantum computing. Here, we propose and realize an all-microwave parametric Controlled-Z (CZ) gates by coupling strength modulation in a superconducting Transmon qubit system with tunable couplers. After optimizing the design of the tunable coupler together with the control pulse numerically, we experimentally realized a 100 ns CZ gate with high fidelity of 99.38\%$ \pm$0.34\% and the control error being 0.1\%. We note that our CZ gates are not affected by pulse distortion and do not need pulse correction, {providing a solution for the real-time pulse generation in a dynamic quantum feedback circuit}. With the expectation of utilizing our all-microwave control scheme to reduce the number of control lines through frequency multiplexing in the future, our scheme draws a blueprint for the high-integrable quantum hardware design. 
		
	\end{abstract}
	
	\maketitle

	Digital quantum gates are the basis for the construction of gate-based quantum error correction, which is proposed to realize universal quantum computing~\cite{universal-review, fowler2012surface}. In practical realization, single-qubit quantum gates are usually easy to implement and have high fidelities~\cite{Single_qubit}. The multi-qubit quantum gates involve the interaction of multiple qubits, thus are much more difficult to implement, and have become one of the important indicators of whether quantum error correction can be achieved~\cite{barends2014superconducting, fowler2012surface}. Profited by the easy-to-control characteristics of superconducting qubits, two-qubit gates, including controlled-Z (CZ)~\cite{barends2014superconducting,ali2021high,li2019realisation, sung2021realization,chen2014qubit,li2020tunable,collodo2020implementation,xu2020high,ye2021realization, TunableBusGate, sete2021parametric, kosen2021building, ganzhorn2020benchmarking}, CNOT~\cite{sheldon2016procedure,kandala2021demonstration}, and fSim gates~\cite{barends2019diabatic}, with fidelity reaching the threshold of error correction has already been realized. 
	
	In superconducting circuits, most of the implementation of two-qubit gates can be classified into unipolar-pulse-based and microwave-based parametric gates. 
	Benefiting from the shorter gate time, the unipolar-pulse-based CZ gates generally achieves higher fidelity than the microwave-based pulse scheme~\cite{barends2014superconducting,ali2021high,li2019realisation,sung2021realization,chen2014qubit,li2020tunable,collodo2020implementation,xu2020high,ye2021realization,sheldon2016procedure,kandala2021demonstration,sete2021parametric,kosen2021building,ganzhorn2020benchmarking}. However, for most of these CZ gates, a significant part of the control error in experiment comes from the time-domain correlation noise caused by the pulse distortion ~\cite{ali2021high,li2019realisation,sung2021realization,collodo2020implementation,ye2021realization}.
	The solution includes correction of the control pulse~\cite{foxen2018high,rol2020time} and net-zero control pulse design~\cite{ali2021high}. Pulse correction requires not only prior knowledge of the distortion formulation but also a large number of computing resources to generate the waveform in advance ~\cite{barends2014superconducting, rol2020time}, and is thus difficult to realize in {real-time waveform generation architecture}~\cite{fowler2012surface, Andersen2019}. The other solution is to utilize the net-zero control pulse at qubits' sweet point ~\cite{ali2021high}. Such scheme requires that {the qubits must work at their sweet point for efficient interaction}. Therefore, higher requirements are put forward for the accuracy-control of qubit frequency in device fabrication. In contrast, the microwave-based two-qubit gates are realized by applying a full microwave control pulse ~\cite{TunableBusGate, sete2021parametric,sheldon2016procedure,kandala2021demonstration}, which is less affected by pulse distortion. However, those gates usually take a longer time to implement, thus encountering more decoherence.
	
	Here, we provide a microwave-based two-qubit gates scheme implemented on a superconducting Transmon qubits architecture with tunable couplers. 
	The CZ gate is realized by applying a net-zero microwave pulse on the coupler flux.
	Compared with microwave-based gates like cross-resonance CNOT gate and other parametric CZ gates, we utilize a wide-range tunable coupler with the coupling strength that can be turned on to be above 10 MHz in both directions.  
	This design significantly shortens the implementation time of microwave CZ gates. 
	With this scheme, we achieved a 100 ns CZ gate with a fidelity of 99.38\%$\pm$0.34\%, where the confidence interval is 95\%. 
	Meanwhile, as the coupler has a coupling off point, we can set the coupler flux for idle and single-qubit gates at the decoupling point. 
	As the qubits are decoupled during idle and single-qubit gates, it is not necessary to consider the effect of residual coupling to detune the frequencies of any qubits far away.
	Therefore, as the single-qubit gate and CZ gate do not contain any unipolar pulse, there is no need to consider pulse correction. 
	We also found that when the frequency of microwaves deviates from the optimal frequency by 5-10MHz, the target quantum state will no longer have a significant evolution. Therefore, we believe that this two-qubit-gate scheme has the potential for frequency multiplexing in the future, especially in combination with suitable on-chip filter designs. 
	
	
	\begin{figure*}
		\centering
		\includegraphics[width=0.8\textwidth]{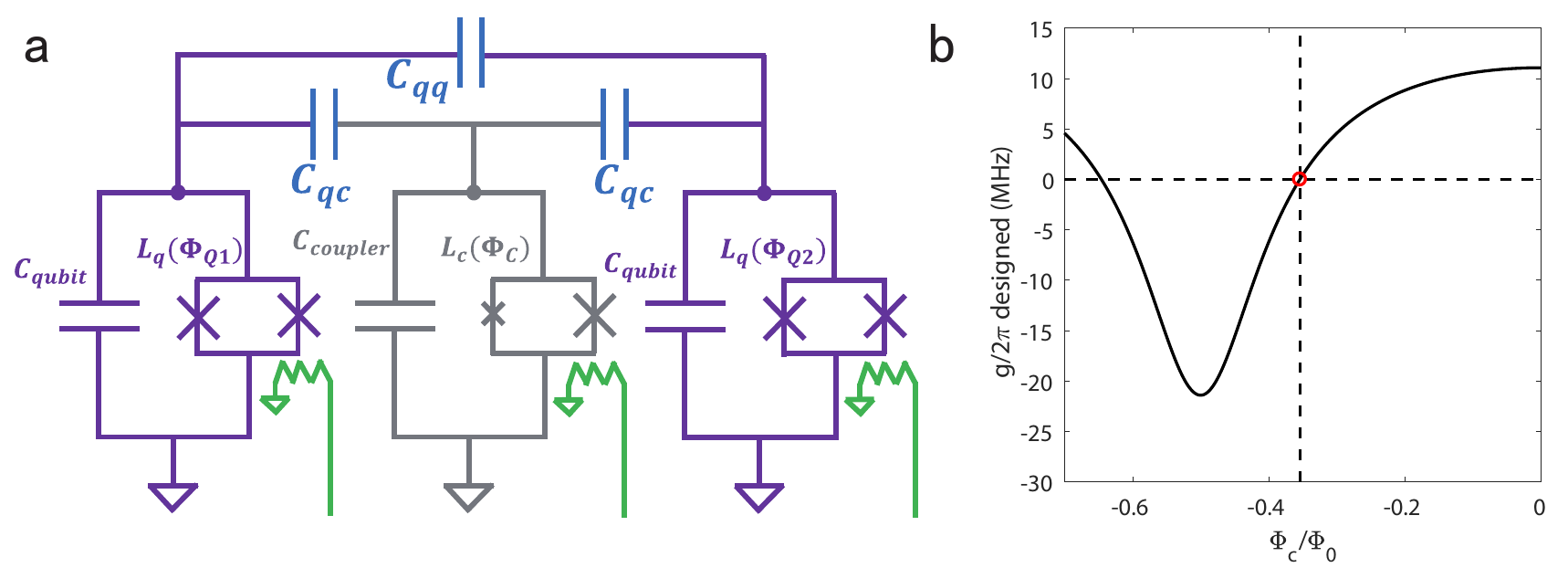}
		\caption{
			(a) Schematic diagram of qubit and coupler circuits. Purple is the qubit, gray is the coupler, blue is the coupling capacitor, and green is the control line. (b) The theoretical design of coupling strength. The capacitive coupler is designed with an asymmetrical junction, and the red circle is the coupling zero point.}
		\label{Figure 1}
	\end{figure*}
	
	The circuit diagram of our quantum coupler qubit design is shown in Fig.\ref{Figure 1}(a). Due to the insufficient precision of the Josephson junction performance prepared in the experiment, we need to fine-tune the qubit frequency according to the decoherence performance and frequency collision. Therefore, our qubit is designed to be frequency tunable with a flux control line, and we can also realize a single-qubit gate by this control line. To reduce the influence of residual XX coupling and    improve the performance of parallel single-qubit gates, our coupler is designed to contain an XX coupling-off point. We set the flux of the coupler near this XX coupling-off point when applying single-qubit gates. 
	
	We designed our coupler parameters based on our simulation results (see supplementary information), specifically $C_{qubit} = 70.9fF$, $C_{coupler} = 80.3fF$, $C_{qq} = 0.71fF$, $C_{qc} = 13.6fF$. The coupler we designed can achieve a coupling strength above 10MHz in both directions to make it efficient to implement a high-fidelity CZ gate. Since the flux pulse is symmetrical during implementing CZ, we design the coupler loop as asymmetric Josephson junctions to avoid the coupler frequency being too close to the qubit frequency.
	For our design parameters, the coupling strength has a minimum value at $\Phi_c/\Phi_0 = 0.5$, measured to be -22MHz, and the frequency difference between the qubit and coupler is larger than 2GHz. 
	
	According to experimental data, we fit the Josephson junction parameters of our qubits and coupler. The $I_c$ of qubits are fitted to be 14.0 nA and 14.0nA and the $I_c$ of coupler are fitted to be 35.0 nA and 70.4 nA. The coupling strength between the two qubits varies from -22MHz to 11MHz (see Fig1 (b)). To trading off the influence of the residual XX coupling and ZZ coupling, we choose the idle working point near the  $g_{xx} = 0$, where $\Phi_c/\Phi_0 = -0.35$, red circle in Fig.\ref{Figure 1}(b). 
	
	\begin{figure*}[hbt]
		\centering
		\includegraphics[width=0.8\textwidth]{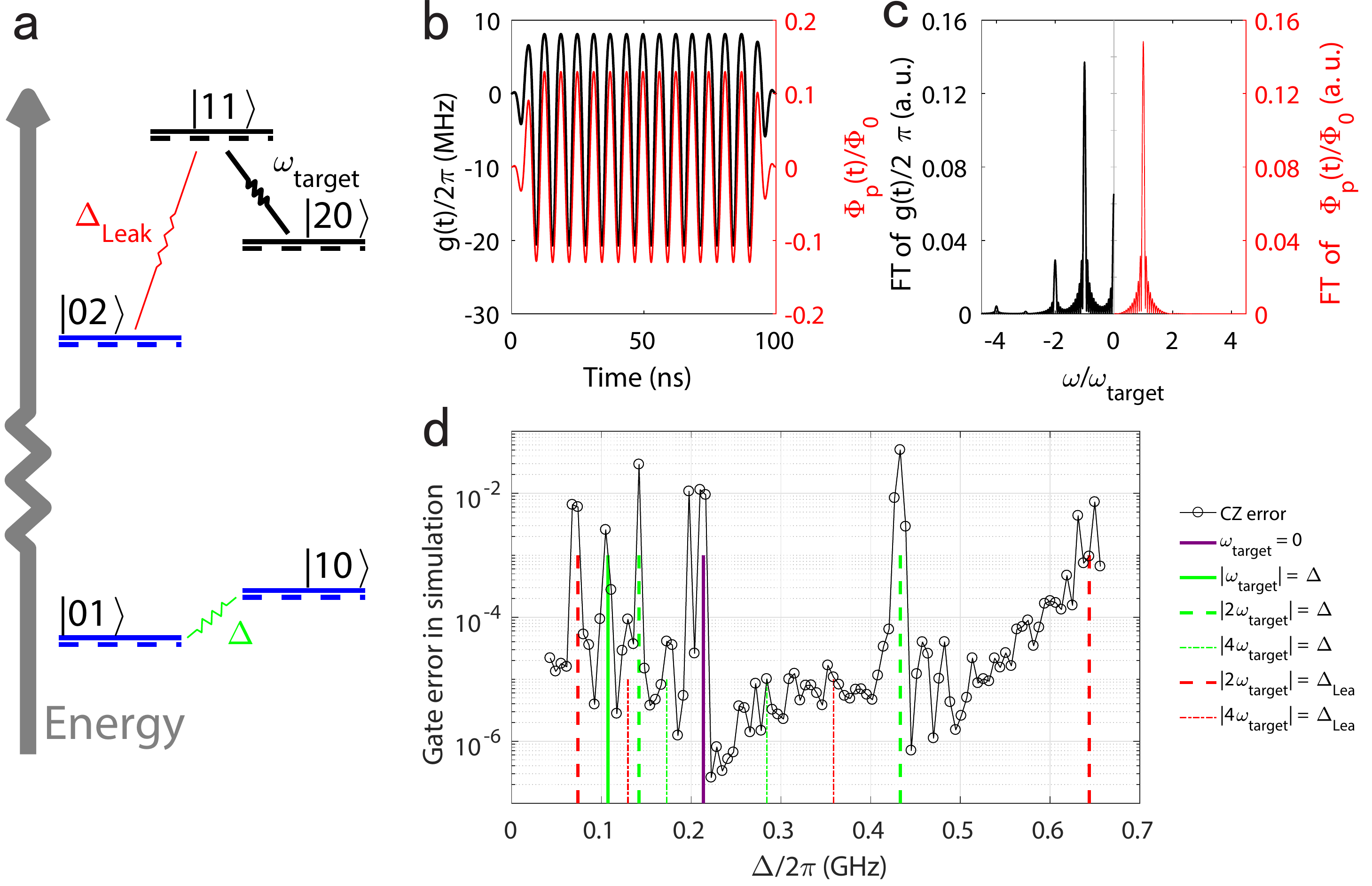}
		\caption{
			(a) Equivalent energy level when microwave pulse is applied to the coupler loop. The solid lines represent the energies at idle time, and the dashed lines represent the energies shift during the input microwave. There are three interactions involved: between $|01\rangle$, $|10\rangle$, between $|11\rangle$, $|02\rangle$, and between $|11\rangle$, $|20\rangle$. (b) The flux in the coupler and the coupling strength between two qubits during the microwave pulse. The relationship between the flux variation and time is a sinusoidal pulse (red solid line), $\Phi_{p}(t) = A(t)\cos({\omega}_{input}t+\phi)$, here $\Phi_{p}(t)$ is the extra flux superimposed on coupler, and $A(t)$ is the envelope of the microwave. The positive and negative coupling is asymmetrical because of the nonlinear relationship between the flux variation and the coupling strength (solid black line). (c) The Fourier component of flux and coupling strength in (b). The main frequency component of flux is at $\omega_{input}$, shown on the right side. From the left side, we can see the frequency components are significantly place at 0, $\omega_{input}$, 2$\omega_{input}$, 3$\omega_{input}$, 4$\omega_{input}$ and even other higher frequency components. (d) Gate error in simulation. The CZ error after 200 iterations for the envelope and coupler design we used (see supplementary information). The vertical lines of different colors indicate that the resonance interaction of $\Delta$ and $\Delta_{leak}$ have been induced, and a large error has been caused.   }
		
		\label{Figure 2}
	\end{figure*}
	
	In the coupling strength tunable transmon qubit, if we ignore the dress state caused by the interaction between the qubit and coupler, the interaction subspace Hamiltonian of two qubits can be simplified as
	\begin{equation}
		\begin{aligned}
			\hat{H}_{sub}(t) = & \hbar\omega_{10}|10\rangle\langle10|+\hbar\omega_{01}|01\rangle\langle01|
			+\hbar\omega_{11}|11\rangle\langle11|+\hbar\omega_{02}|02\rangle\langle02|
			+\hbar\omega_{20}|20\rangle\langle20|\\
			&+{\hbar}g(t)(|10\rangle\langle01|+|01\rangle\langle10|)
			+{\sqrt{2}\hbar}g(t)(|02\rangle\langle11|+|11\rangle\langle02|)\\
			&+{\sqrt{2}\hbar}g(t)(|20\rangle\langle11|+|11\rangle\langle20|).
		\end{aligned}
		\label{eq1}
	\end{equation}
	
	We assume the difference of qubit frequency as $\Delta = \omega_{10}-\omega_{01} \textgreater 0$, and coupling strength as $g(t) = \Omega(t)\cos({\omega}_{target}t+\phi)$. If $\omega$ satisfies that $\omega_{target} = |\omega_{11}-\omega_{20}|$, the $|11\rangle$ and $|20\rangle$ will interact in resonance, as shown by the black line in Fig.\ref{Figure 2}(a). Here $\Omega(t)$ is the envelope function and $\omega_{target}$ is the driven frequency. To set a reasonable working point, it must satisfy that $|{\Delta_{leak}-\omega_{target}}|  \gg \Omega(t)$ and $|\Delta-\omega_{target}| \gg \Omega(t)$ to avoid the other two interactions, as shown by the thin red line in Fig.\ref{Figure 2}(a), here $\Delta_{leak} = |\omega_{11}-\omega_{02}|$.
	
	However, the above Hamiltonian is only an approximate demonstration. For the actual tunable coupling qubits, one needs to be cautious about the following two points. First, the energy eigenstate is a dressed state, and $\omega_{01}$ changes with coupler flux. The qubit center frequency will have a shift of 0-5MHz according to the driving amplitude $\Omega(t)$, which will impact the decoherence performance of the qubit. Second, the relationship between coupling strength and coupler flux is nonlinear. To generate a net-zero microwave pulse, the variation in flux is sinusoidal-like, see red line in Fig.\ref{Figure 2}(b), and the symmetrical flux variation will produce asymmetrical coupling strength variation, black line in Fig.\ref{Figure 2}(b). This means that a monochromatic flux pulse will result in a coupling strength pulse with multiple peaks in the frequency domain, see Fig.\ref{Figure 2}(c). Therefore, we should also consider the influence of these frequency components when selecting the qubit working point, see Fig.\ref{Figure 2}(d). We checked the gate error of the CZ gate that can be achieved under different frequency settings $\Delta$ after 200 iterations of Nelder Mead algorithm~\cite{nelder1965simplex,mckinnon1998convergence,rol2017restless} by numerical method. We found that if the unwanted resonance interactions induced by the $\omega_{target}$ and $2\omega_{target}$ are avoided, a CZ gate with fidelity higher than 99.99\% can be easily achieved.

	The two qubits we used to perform the experiment are named Q1 and Q2. Considered the decoherent performance and avoiding unwanted resonance interactions, we set the qubit idle point shown in Tab.\ref{tab1}. The performances in Tab.\ref{tab1} are measured at $\Phi_c/\Phi_0 = -0.35$, and $T_{1,CZ}$ is obtained by averaging the frequency from 0 to 5MHz below the idle point. 
	
	\begin{table}[hbt]
		\centering
		\caption{Qubits performances at selected working points.}
		\label{tab1}
		\setlength\tabcolsep{10pt}
		\newcommand{\tabincell}[2]{\begin{tabular}{@{}#1@{}}#2\end{tabular}}
		\begin{tabular}{ccccccc}
			\hline
			\hline
			qubit & $f_{01, max}$ & $f_{01, idle}$ & anharmonicity & $T_{1, idle}$ & $T_{1, CZ}$ & $T_{2,idle}^*$\\  \hline
			Q1  &4.770 GHz  &4.770 GHz  &-232 MHz  &32.2${\mu}s$  &36.0 ${\mu}s$  & 13.5 ${\mu}s$ \\  \hline
			Q2  &4.847 GHz  &4.839 GHz  &-230 MHz  &58.5${\mu}s$  &45.6 ${\mu}s$  & 13.3 ${\mu}s$ \\ \hline
			\hline
		\end{tabular}
	\end{table}
	
	\begin{figure*}[hbt]
		\centering
		\includegraphics[width=0.8\textwidth]{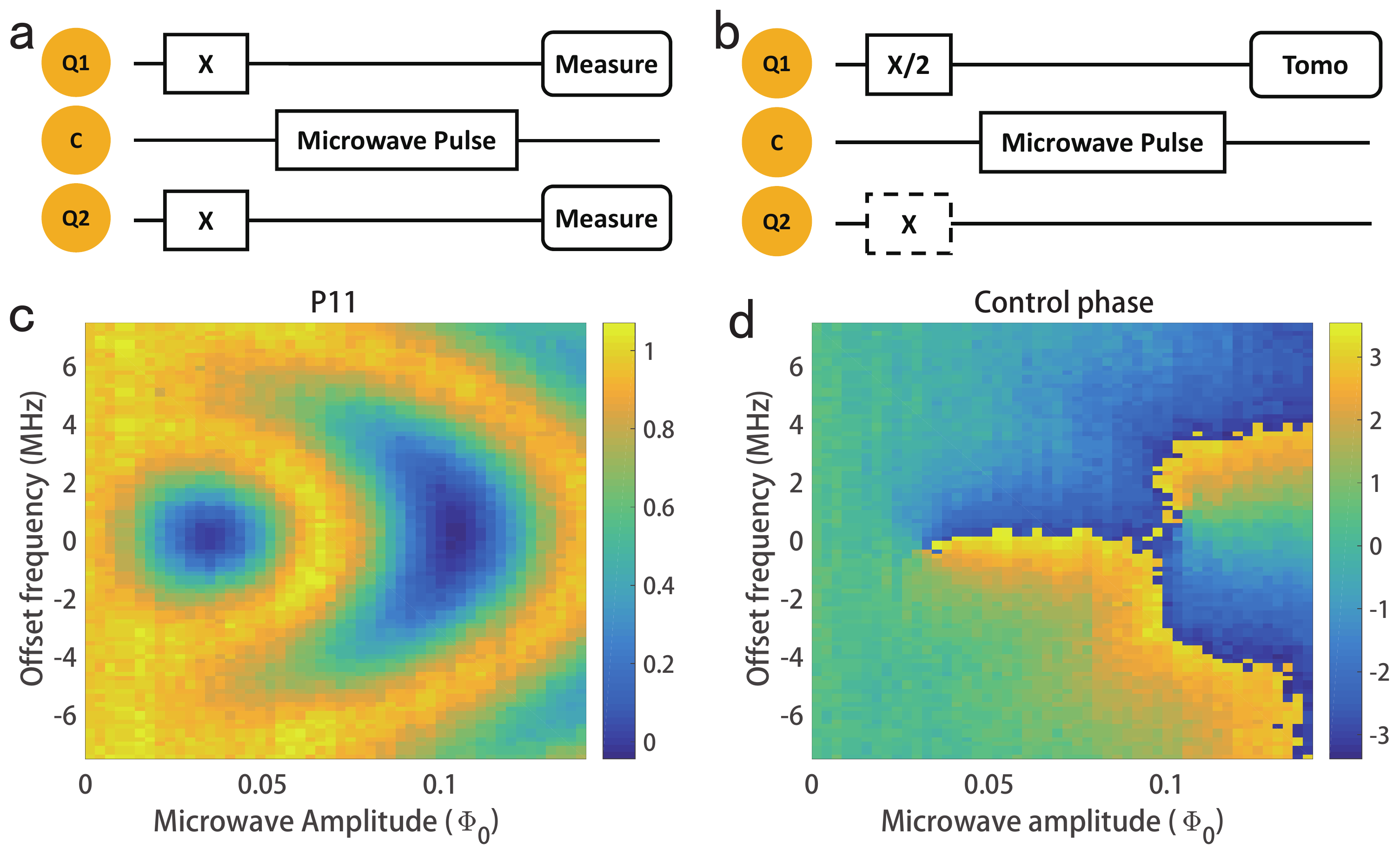}
		\caption{
			(a) Quantum sequence for detecting the swap between $|11\rangle$ and $|20\rangle$. We prepare two qubits in $|11\rangle$ by two X gates for each qubit, then apply microwave pulse at coupler control line and measure the final population of $|11\rangle$. (b) Quantum sequence for detecting the control phase. We prepare Q1 in a superposition state and measure the phase $\theta_1$ and $\theta_2$ of Q1 after microwave pulse according to applying X gate to Q2 or not. We calculate the control phase according to the formula $\theta_{control}$ = $\theta_1-\theta_2$. (c) The result is measured in the experiment by the sequence in (a). The x-axis is the averaging amplitude $\bar{A}(t)$ of the microwave pulse during the active time, and the y-axis is the deviation of the microwave pulse frequency from the resonance frequency. (d) The result is measured in the experiment by the sequence in (b). The control phase is measured by Q1. }
		\label{Figure 3}
	\end{figure*}
	
\begin{figure*}[hbt]
		\centering
		\includegraphics[width=0.8\textwidth]{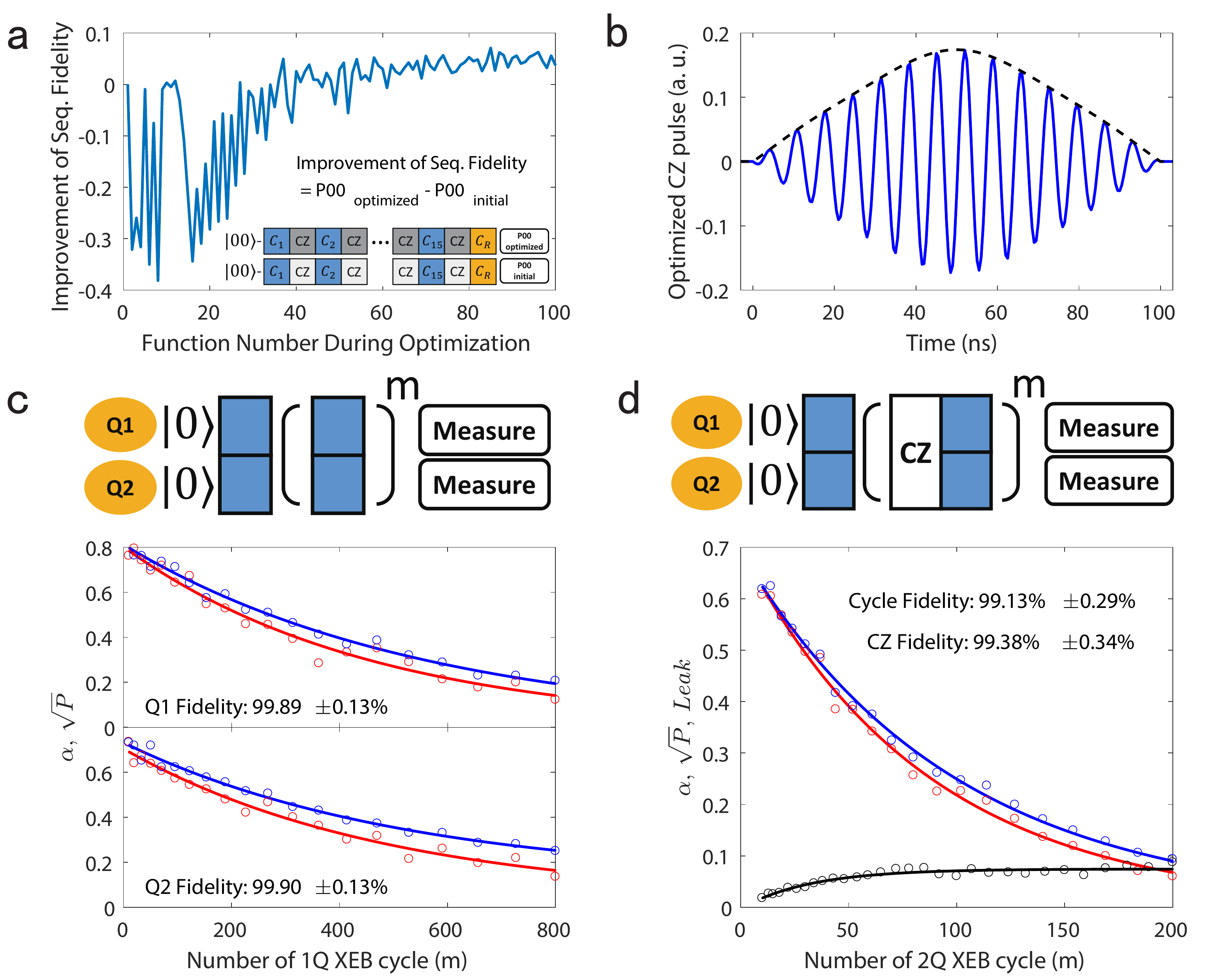}
		\caption{
			(a) Optimization of CZ gate. The cost function is obtained by sequence fidelity of interleaved RB that contains 15 Clifford gates and an inversed gate. (b) The optimized control microwave pulse. The solid line is the applied microwave, and the dashed line is the envelope. (c) XEB and SPB results of single-qubit cycle. The sequence of the single-qubit cycle is shown in the upper part. The circle and the solid lines are experimental data and the fitting result of Q1 and Q2. Blue and red are data of $\sqrt{P}$ and $\alpha$. (d) XEB, SPB, leakage results of CZ cycle. The sequence of the CZ cycle is shown in the upper part. Blue, red and black are data of $\sqrt{P}$, $\alpha$, and leakage extracted by the CZ cycle. }
		\label{Figure 4}
		
	\end{figure*}

	We set the microwave active time to 250 ns and select a suitable envelope $A(t)$ to measure the population exchange between $|11\rangle$ and $|20\rangle$ and the control phase under different microwave amplitudes. Fig.\ref{Figure 3}(a) depicts the circuit for measuring transition rate from and back to $|11\rangle$ state where Fig.\ref{Figure 3}(c) describes this quantity as a function of microwave amplitude $A(t)$ and frequency. Fig.\ref{Figure 3}(b) shows the circuit for measuring relative phase between $|0\rangle$ and $|1\rangle$ state with Q2 being either $|0\rangle$ or $|1\rangle$. The difference between the two outcomes relates to the control phase, see Fig.\ref{Figure 4}(d). We found that when the microwave amplitude is near 0.06, and the microwave frequency is near the resonance frequency, the population of $|11\rangle$ in the final state approaches 1, and the control phase is close to $\pi$, which implies a reliable CZ gate. Such results serve as a rough estimation of the parameters of a CZ gate, from which we will fine-tune each value by considering more realistic situations.
	
	The error of the CZ gate in the experiment mainly includes the following four types: decoherence error, leakage error caused by residue population in $|20\rangle$ and $|02\rangle$, phase error caused by inaccurate control phase, and swap error caused by interaction between $|01\rangle$ and  $|10\rangle$~\cite{sendelbach2009complex,fried2019assessing,megrant2012planar}. 
	To mitigate these errors, we need to optimize the envelope of the control pulse. We first parameterize the envelope as
	
	\begin{equation}
		\begin{aligned}
			A(t) = \lambda_{1}\sin{\frac{{\pi}t}{T_a}}+\lambda_{2}(1-\cos\frac{2{\pi}t}{T_a})+\lambda_{3}\sin{\frac{3{\pi}t}{T_a}}+\lambda_{4}(1-\cos\frac{4{\pi}t}{T_a}).
		\end{aligned}
		\label{eq2}
	\end{equation}
	
	Here $T_a$ is the microwave active time, and $\vec{\lambda}$ is the envelope parameter that needs to be optimized. To reduce the decoherence error, we set $T_a$ = 100 ns. Furthermore, we initialized the $\vec{\lambda}$ ratio according to the result of the numerical simulation ($\lambda_1:\lambda_2:\lambda_3:\lambda_4 = -0.0760:1.0000:0.4222:-0.1636$), then we measured the swap probability and control phase to determine the initial value of $\bar{A}(t)$ and microwave frequency (see Fig.\ref{Figure 3}). Next, we optimized $\lambda_1$, $\lambda_3$, $\lambda_4$, $\bar{A}(t)$ and microwave frequency to improve the fidelity of the CZ gate through 50 to 100 iterations, as described in Fig.\ref{Figure 4}(a). The evaluation function related to CZ fidelity is measured by the randomized benchmarking (RB) method ~\cite{knill2008randomized,epstein2014investigating,proctor2017randomized}. Some experimental methods can reduce the noise in the measurement of this evaluation function. Depending on the error tolerance, each measurement may take 20 to 60 seconds, which amounts to a total cost of about 1 to 2 hours for a complete optimization, including data processing. 
	
	After 100 iterations, we obtain the optimized control microwave pulse of the CZ gate as shown in Fig.\ref{Figure 4}(b). In addition to the active time of 100 ns, we add three ns idle time before and after the active time to avoid timing misalignment and microwave residue. The average amplitude is determined by optimization, and the actual CZ pulse reaching the coupler will be smaller compared to the same input amplitude of DC bias due to attenuation and bonding-wire reflections at $\omega_{target}$.
	
	We use cross-entropy benchmarking (XEB) and speckle purity benchmarking (SPB) ~\cite{boixo2018characterizing,barends2019diabatic,arute2019quantum} to quantify the error sourse of our single-qubit gate and CZ gate. The XEB circuits and results of single-qubit gate and CZ gate are shown in Fig.\ref{Figure 4}(c) and (d). We fitted the sequence fidelity and purity fidelity data according to $\alpha = Ap_{xeb}^m+B$, $\sqrt{P} = Ap_{spb}^m+B$~, the formulas of calculation $\alpha$ and $\sqrt{P}$ are described in supplementary information, and we calculated the cycle Pauli error from the formula $r_{p,xeb(spb)} = (1-p_{xeb(spb)})(4^N-1)/4^N$, here N is the number of qubit. The Pauli error of CZ gate is extracted from formula $(1-r_{p, cycle}) = (1-r_{p,CZ})(1-r_{p,Q1})(1-r_{p,Q2})$. The population of leakage is fitted by formula of $\epsilon_{leak} = Ap_{leak}^m+B$, and we calculated leakage error such that $r_{leak} = -A(1-p_{leak})(4^N-1)/4^N$. We extract control Pauli error as $r_{p,ctrl} = r_{p,xeb}-r_{p,spb}$ and decoherence Pauli error as $r_{p,dec} = r_{p,spb}-r_{leak}$. The gate fidelity can be obtained as $F = 1-r_p2^N/(2^N+1)$, and CZ fidelity calculated to be $99.38\%\pm0.34\%$ with 95\% confidence. The single-qubit gate and CZ gate error budget is shown in Tab.\ref{table2}. 
	
	\begin{table}[hbt]
		\centering
		\caption{Error budget of single-qubit and CZ gates.}
		\setlength\tabcolsep{5pt}
		\newcommand{\tabincell}[2]{\begin{tabular}{@{}#1@{}}#2\end{tabular}}
		\begin{tabular}{cccccccccc}
			\hline
			\hline
			Gate & T (ns) &$p_{xeb}$ & $r_{p,xeb}$ & $p_{spb}$ & $r_{p,spb}$ & $r_{leak}$ & $r_{p,dec}$ &$r_{p,ctrl}$ & Fidelity\\  \hline
			Q1-$\pi/2$ & 50 & 99.78\%  &0.16\%  &99.82\%  & 0.13\% & \textless0.01\% &  0.12-0.13\% &0.03\%  & 99.89\% \\ 
			\hline
			Q2-$\pi/2$ & 50 & 99.80\%  &0.15\%  &99.81\%  & 0.14\% & \textless0.01\% & 0.13-0.14\% &0.01\%  & 99.90\% \\ 
			\hline
			cycle-CZ &156 & 98.84\% & 1.09\% & 98.99\% & 0.95\% & 0.22\% & 0.73\% & 0.14\% & 99.13\%\\
			\hline
			CZ & 106 & &0.78\% & & 0.68\%  & 0.20-0.22\% & 0.46-0.48\% & 0.10\% & 99.38\%\\
			\hline
			\hline
		\end{tabular}
		\label{table2}
	\end{table}
	
	For the numerical simulation without considering decoherence, the error of the CZ gate comes from leakage error, phase error, and swap error. 
	The total error of these three parts can be suppressed to less than 0.01\% with a duration of 100ns easily. Limited by the fluctuation of the readout, we cannot sufficiently optimize the parameters of the control pulse, leading to the control error and leakage error (in Tab.\ref{table2}) higher than numerical results. With the improvement of reading performance and system stability, it is hopeful of reducing the experiment error further. 
	
	In summary, we have demonstrated a fast and high-fidelity all-microwave CZ gate scheme on a new coupler structure design.
	The two-qubit gates can be realized fast thanks to this specific coupler design with stronger coupling strength. Meanwhile, all qubits are decoupled except for applying two-qubit gates, facilitating parallel optimization on large-scale quantum devices in the future. Also, our CZ gate scheme is designed without any pulse correction, thus can be easily applied in real-time pulse generation~\cite{fowler2012surface}. 
	Another essential advantage of our gate scheme is that it can be combined with frequency multiplexing technology. 
	Microwave control pulses combined with on-chip filters~\cite{dai2014silicon,kobe2017review} can be a feasible solution to reduce the number of control lines, as we can deliver the microwave of different frequencies through a common signal input to corresponding couplers through on-chip filters. Considering all these advantages above, our all-microwave CZ gate scheme can have broad application prospects in large-scale quantum hardware design in the future. 
	
	\begin{acknowledgments}
		
		The authors thank the USTC Center for Micro- and Nanoscale Research and Fabrication for supporting the sample fabrication and QuantumCTek Co., Ltd. for supporting the fabrication and the maintenance of room-temperature electronics.
		\textbf{Funding:}
		This research was supported by the National Key R\&D Program of China, Grant 2017YFA0304300, the Chinese Academy of Sciences, Anhui Initiative in Quantum Information Technologies, Technology Committee of Shanghai Municipality, National Science Foundation of China (Grants No. 11905217),  Natural Science Foundation of Shanghai (Grant No. 19ZR1462700), and Key-Area Research and Development Program of Guangdong Provice (Grant No.2020B0303030001). National Natural Science Foundation of China (Grants No. 11905294), China Postdoctoral Science Foundation. 
		
	\end{acknowledgments}
	
	%

\end{document}